\pdfoutput=1
\documentclass{svjour3}
\journalname{Few-Body Systems}
\usepackage{graphicx}
\usepackage{amsmath}
\renewcommand{\vec}{\mathbf}
\begin{document}
\title{A nuclear model with explicit mesons}
\author{D.V.~Fedorov}
\institute{D.V.~Fedorov \at
	Aarhus University, Aarhus, Denmark\\
	\email{fedorov@phys.au.dk}
}
\date{}
\maketitle
	\begin{abstract}

A nuclear model is proposed where the nucleons interact by emitting and
absorbing mesons, and where the mesons are treated explicitly.  A nucleus
in this model finds itself in a quantum superposition of states with
different number of mesons.  Transitions between these states hold the
nucleus together.

The model---in its simplest incarnation---is applied to the deuteron,
where the latter becomes a superposition of a neutron-proton state and a
neutron-proton-meson state.  Coupling between these states leads to an
effective attraction between the nucleons and results in a bound state
with negative energy, the deuteron.  The model is able to reproduce
the accepted values for the binding energy and the charge radius of
the deuteron.

The model, should it work in practice, has several potential advantages
over the existing non-relativistic few-body nuclear models: the reduced
number of model parameters, natural inclusion of few-body forces, and
natural inclusion of mesonic physics.
	\end{abstract}

\section{Introduction}

In the low-energy regime the nucleons are believed to interact by
exchanging mesons~\cite{yukawa,jensen,langanke}. However the accepted
contemporary non-relativistic few-body nuclear models customarily
eliminate mesons from the picture and introduce instead phenomenological
meson-exchange-inspired nucleon-nucleon potentials tuned to reproduce
available experimental data~\cite{langanke,bonn,av18,historical}.

In this contribution the meson-exchange paradigm is going to be applied
literally by allowing nucleons to explicitly emit and absorb mesons
which will be treated on the same footing as the nucleons.  A nucleus
in this model will be a superposition of states with different number
of emitted mesons.

Since it takes energy to generate a meson, in the low-energy regime the
states with mesons will find themselves under a potential barrier equal
to the total mass of the mesons.  One might expect that in the first
approximation only the state with one meson will contribute significantly.

In one-meson approximation a nucleus becomes a superposition of two
subsystems: a subsystem with zero mesons, and a subsystem with one meson.
The corresponding Hamiltonian is then given as a matrix,
       \begin{equation}\label{eq-1}
H=\left(
\begin{array}{cc}
K_N & W  \\
W & K_N+K_\sigma+m_\sigma \\
\end{array}
\right)
\;,
       \end{equation}
where $K_N$ is the kinetic energy of nucleons, $K_\sigma$ is the kinetic
energy of the meson, $m_\sigma$ is the mass of the meson, and $W$ is the
operator that couples these two subsystems by generating/annihilating
the meson.
The corresponding Schrodinger equation for the nucleus is then given as
       \begin{equation}
\left(
\begin{array}{cc}
K_N & W  \\
W & K_N+K_\sigma+m_\sigma \\
\end{array}
\right)
\left(
\begin{array}{c}
\psi_N \\
\psi_{\sigma N}
\end{array}
\right)
=E
\left(
\begin{array}{c}
\psi_N \\
\psi_{\sigma N}
\end{array}
\right)
\;,
       \end{equation}
where $\psi_N$ is the wave-function of the subsystem with nucleons only;
$\psi_{\sigma N}$ is the wave-function of the subsystem with nucleons
and a meson; and $E$ is the energy.  If $E<m_\sigma$ the meson is under
barrier and cannot leave the nucleus.

The potential advantages of this model, should it work in practice, are
the reduced number of parameters (possibly only few parameters per meson);
natural inclusion of few-body forces (expected to arise from few-meson
states); and natural inclusion of mesonic physics.  One disadvantage is
a substantially increased computational load: an $N$~body problem becomes
(at least) a coupled $N$ plus $N+1$~body problem.

The lightest mesons are pions and they should presumably be
included first.  However in this first conceptual trial of the
model we shall for simplicity only include the scalar-isoscalar
sigma-meson~\cite{sigma-meson} which is assumed to be responsible for the
bulk of the intermediate-range attraction in nuclei~\cite{historical}
and which is also the lightest meson in relativistic mean-field
theories~\cite{manka}.

In the next sections we shall
i) describe the practical application of the model to the deuteron;
ii) describe the correlated Gaussian method for coupled few-body systems
to be used to solve the problem numerically;
iii) present the results for the deuteron;
and iv) give a conclusion.  Several relevant formulae are collected in
the Appendix.

\section{Deuteron as a system of two nucleons and a meson}

The deuteron in this model consists of two coupled subsystems: a two-body
neutron-proton subsystem and a three-body neutron-proton-meson subsystem.
The wave-function $\psi$ of the deuteron is then a two-component
structure,
       \begin{equation}
\psi=\left(
\begin{array}{c}
\psi_\mathrm{np}(\vec r_\mathrm{n},\vec r_\mathrm{p})\\
\psi_\mathrm{\sigma np}(\vec r_\mathrm{\sigma},\vec r_\mathrm{n},\vec r_\mathrm{p})
\end{array}
\right)
\;,
       \end{equation}
where $\psi_\mathrm{np}$ is the wave-function of the two-body subsystem;
$\psi_\mathrm{\sigma np}$ is the wave-function of the tree-body
subsystem; and where
$\vec r_\mathrm{n}$,
$\vec r_\mathrm{p}$,
$\vec r_\mathrm{\sigma}$ and are the coordinates
of the neutron, the proton, and the sigma-meson.

Correspondingly, the Hamiltonian is a matrix,
	\begin{equation}
H=\left(
\begin{array}{cc}
K_\mathrm{n}+K_\mathrm{p} & W \\
W       & K_\mathrm{n}+K_\mathrm{p}+K_\mathrm{\sigma}+m_\sigma
\end{array}
\right)
\;,
	\end{equation}
where $K_\mathrm{n}$, $K_\mathrm{p}$, $K_\mathrm{\sigma}$ are the kinetic
energy operators for the neutron, the proton, and the sigma-meson; $W$ is
the coupling operator (to be introduced later); and $m_\mathrm{\sigma}$
is the mass of the sigma-meson.

It is of advantage to introduce relative Jacobi coordinates,
       \begin{equation}
\vec r_\mathrm{np} = \vec r_\mathrm{p}-\vec r_\mathrm{n}
\;,\;\;
\vec r_\mathrm{\sigma np} =
\frac{
m_\mathrm{n}\vec r_\mathrm{n}
+m_\mathrm{p}\vec r_\mathrm{p}
}{m_\mathrm{n}+m_\mathrm{p}}
-\vec r_\mathrm{\sigma}
\;,
       \end{equation}
with the corresponding relative kinetic energy operators,
       \begin{equation}
K_\mathrm{np}=-\frac{\hbar^2}{2\mu_\mathrm{np}}
\frac{\partial^2}{\partial\vec r_\mathrm{np}^2}
\;,\;\;
K_\mathrm{\sigma np}=-\frac{\hbar^2}{2\mu_\mathrm{\sigma np}}
\frac{\partial^2}{\partial\vec r_\mathrm{\sigma np}^2}
\;,
       \end{equation}
were the reduced masses are given as
       \begin{equation}
\mu_\mathrm{np}=
\frac{m_\mathrm{n}m_\mathrm{p}}{m_\mathrm{n}+m_\mathrm{p}}
\;,\;\;
\mu_\mathrm{\sigma np}=
\frac{m_\mathrm{\sigma}(m_\mathrm{n}+m_\mathrm{p})}
{m_\mathrm{\sigma}+m_\mathrm{n}+m_\mathrm{p}}
\;.
       \end{equation}
In the center-of-mass system the Hamiltonian is then given as
	\begin{equation}
H=
\begin{pmatrix}
K_\mathrm{np} & W \\
W       & K_\mathrm{np}+K_\mathrm{\sigma np}+m_\sigma
\end{pmatrix}
\;,
	\end{equation}
with the corresponding Schrodinger equation,
	\begin{equation}\label{eq-d-sch}
\begin{pmatrix}
K_\mathrm{np} & W \\
W       & K_\mathrm{np}+K_\mathrm{\sigma np}+m_\sigma
\end{pmatrix}
\begin{pmatrix}
\psi_\mathrm{np}(\vec r_\mathrm{np}) \\
\psi_{\sigma pn}(\vec r_\mathrm{np},\vec r_\mathrm{\sigma np})
\end{pmatrix}
= E
\begin{pmatrix}
\psi_{pn}(\vec r_\mathrm{np}) \\
\psi_{\sigma pn}(\vec r_\mathrm{np},\vec r_\mathrm{\sigma np})
\end{pmatrix}
\;.
	\end{equation}

The coupling term for a scalar meson can be introduced in the integral
form as
	\begin{equation}
\langle\psi_\mathrm{np}\bigm| W \bigm| \psi_\mathrm{\sigma np}\rangle
=\int d^3r_\mathrm{np} d^3r_\mathrm{\sigma np}
\psi_{np}^*(\vec r_\mathrm{np})
W(\vec r_\mathrm{np},\vec r_\mathrm{\sigma np})
\psi_\mathrm{\sigma np}(\vec r_\mathrm{np},\vec r_\mathrm{\sigma np})
\;.
	\end{equation}
One can assume that the nucleons themselves are already ``dressed'' with
mesons and that the $W$-operator only accounts for the
``extra'' mesons generated due to the presence of another nucleon. The
kernel $W(\vec r_\mathrm{np},\vec r_\mathrm{\sigma np})$ then has to be
of short range and has to vanish when either of the three particles is
at larger distances.
One of the simplest form is a Gaussian,
	\begin{equation}\label{eq-w-gau}
W(\vec r_\mathrm{np},\vec r_\mathrm{\sigma np})=
S_\sigma\exp \left(-\frac{\vec
r_\mathrm{np}^2+\vec r_\mathrm{\sigma np}^2}{b_\sigma^2}\right)
\;,
	\end{equation}
where the strength $S_\sigma$ and the range $b_\sigma$ are the parameters of
the model.

The coupled Schrodinger equation~(\ref{eq-d-sch}) can be solved
numerically using the correlated Gaussian method
described in the next section.

\section{Correlated Gaussian method for coupled few-body systems}

In the correlated Gaussian method~\cite{suzuki-varga} the wave-function
of a few-body quantum system is represented as a linear combination of
correlated Gaussians,
       \begin{equation}
e^{-\vec r^\mathrm T A\vec r}
\equiv
\exp\left(-\sum_{i,j=1}^n A_{ij}\vec r_i\cdot\vec r_j\right)
\;,
       \end{equation}
where
$\vec r=\begin{pmatrix}\vec r_1 &\vec r_2 &\dots &\vec r_n\end{pmatrix}^\mathrm{T}$
is a
column-vector
size-$n$
set of the coordinates of the system; $A$ is the matrix of parameters;
and $\vec r_i\cdot\vec r_j$ signifies the scalar product of the two
coordinates.

For the three-body $\sigma$np-subsystem the set $\vec r^{(\sigma)}$
of the center-of-mass coordinates is a two-component structure,
       \begin{equation}
\vec r^{(\sigma)}= \left(
\begin{array}{c} \vec r_\mathrm{np}\\
\vec r_\mathrm{\sigma np}
\end{array}
\right)
\;,
       \end{equation}
and the corresponding matrix $A_\sigma$ is a two-times-two matrix. For the
two-body np-subsystem the set $\vec r^{(\mathrm{d})}$ of the center-of-mass
coordinates is a simply a one-component structure,
       \begin{equation}
\vec r^{(\mathrm{d})}= \left(
\begin{array}{c} \vec r_\mathrm{np}
\end{array}
\right)
\;,
       \end{equation}
and the corresponding matrix $A^{(\mathrm{d})}$ is a one-times-one
matrix. Despite the unit dimension we shall keep the matrix notation
for consistency.

Now the wave-functions of the two subsystems are represented as
       \begin{eqnarray}\label{gau}
\psi_\mathrm{np}(\vec r^{(\mathrm{d})})&=&
\sum_{i=1}^{n^{(\mathrm{d})}}
c^{(\mathrm d)}_i
e^{-\vec r^{(\mathrm{d})\mathrm T}A^{(\mathrm d)}_i\vec r^{(\mathrm{d})}}
\equiv
\sum_{i=1}^{n^{(\mathrm{d})}}
c^{(\mathrm d)}_i
\langle \vec r^{(\mathrm{d})} \bigm| A^{(\mathrm d)}_i \rangle
\;,\nonumber\\
\psi_\mathrm{\sigma np}(\vec r^{(\mathrm{\sigma})})
&=&
\sum_{j=1}^{n^{(\sigma)}}
c^{(\mathrm \sigma)}_j
e^{-\vec r^{(\sigma)\mathrm T}A^{(\sigma)}_j\vec r^{(\sigma)}}
\equiv
\sum_{j=1}^{n^{(\sigma)}}
c^{(\mathrm \sigma)}_j
\langle \vec r^{(\sigma)} \bigm| A^{(\mathrm \sigma)}_j \rangle
\;,
       \end{eqnarray}
where $n^{(\mathrm d)}$ and $n^{(\sigma)}$ are the number of Gaussians
in the subsystems, and where
$$\{c^{(\mathrm d)}_i,c^{(\sigma)}_j,A^{(\mathrm d)}_i,A^{(\sigma)}_j\}$$
are variational parameters.  The non-linear parameters
$\{A^{(\mathrm d)}_i,A^{(\sigma)}_j\}$
are chosen stochastically while the linear
parameters $c\doteq\{c^{(\mathrm d)}_i,c^{(\sigma)}_j\}$ are obtained
by solving the generalized eigenvalue problem,
       \begin{equation}\label{gensym}
\mathcal H c = E \mathcal N c \;,
       \end{equation}
where the matrices $\mathcal N$ and $\mathcal H$ are the overlap and
the Hamiltonian matrix in the Gaussian representation~(\ref{gau}).
The matrices have the following two-times-two block structure,
       \begin{equation}
\mathcal H=
\begin{pmatrix}
\left\langle A^{(\mathrm d)}_i
\bigm| K_\mathrm{np} \bigm|
A^{(\mathrm d)}_{i'} \right\rangle
& \left\langle A^{(\mathrm d)}_i \bigm| W \bigm| A^{(\sigma)}_j \right\rangle \\
\left\langle A^{(\sigma)}_j \bigm| W \bigm| A^{(\mathrm d)}_i \right\rangle &
\left\langle A^{(\sigma)}_j \bigm| K_\mathrm{np}+K_\mathrm{\sigma np} +m_\sigma
\bigm| A^{(\sigma)}_{j'} \right\rangle
\end{pmatrix}
\;,
       \end{equation}
       \begin{equation}
\mathcal N=
\left(
\begin{array}{cc}
\left\langle A^{(\mathrm d)}_i \bigm| A^{(\mathrm d)}_{i'} \right\rangle & 0 \\
0 & \left\langle A^{(\sigma)}_j \bigm| A^{(\sigma)}_{j'} \right\rangle
\end{array}
\right)
\;,
       \end{equation}
where
$i,i'=1,\dots,n^{(\mathrm d)}$ and
$j,j'=1,\dots,n^{(\sigma)}$.

One of the advantages of the correlated Gaussian method is that the
matrix elements of the matrices $\mathcal N$ and $\mathcal H$ are
analytic~\cite{fedorov2017analytic}. The cross terms are zero
for all operators $X$ except for the coupling operator $W$,
       \begin{equation}
\langle A^{(\mathrm d)}_i \bigm| X \bigm| A^{(\sigma)}_j \rangle = 0
\;.
       \end{equation}
The
overlap is given as
       \begin{equation}
\langle A \bigm| A' \rangle =
\left(
\frac{\pi^{\mathrm{size}(A)}}{\det\left(A+A'\right)}
\right)^{\frac32}
\;.
       \end{equation}
The matrix element of the kinetic energy operator is given as
       \begin{equation}
\langle A \bigm|
-\frac{\partial}{\partial\vec r}
K
\frac{\partial}{\partial\vec r^\mathrm T}
\bigm| A' \rangle=
6\; \langle A \bigm| A' \rangle \; \mathrm{trace}\left(AKA'(A+A')^{-1}\right)
\;,
       \end{equation}
where the matrix $K$ for the $\sigma$np-subsystem is given as
       \begin{equation}
K^{(\sigma)}=
\left(
\begin{array}{cc}
\frac{\hbar^2}{2\mu_\mathrm{np}} & 0 \\
0 & \frac{\hbar^2}{2\mu_\mathrm{\sigma np}}
\end{array}
\right)
\;,
       \end{equation}
and for the np-subsystem as
       \begin{equation}
K^{(\mathrm d)}=
\left(
\begin{array}{c}
\frac{\hbar^2}{2\mu_\mathrm{np}} \\
\end{array}
\right)
\;.
       \end{equation}

The matrix element of the coupling operator~(\ref{eq-w-gau}) is given as
       \begin{equation}
\langle A^{(\mathrm d)} \bigm| W \bigm| A^{(\sigma)} \rangle =
S_\sigma \langle \tilde A \bigm| A^{(\sigma)} \rangle
\;,
       \end{equation}
where the matrix $\tilde A$ is given as
       \begin{equation}
\tilde A=
\left(
\begin{array}{cc}
A^{(\mathrm d)}+\frac1{b_\sigma^2} & 0 \\
0 & \frac1{b_\sigma^2}
\end{array}
\right)
\;.
       \end{equation}

The generalized eigenvalue problem~(\ref{gensym}) is solved using
the standard algorithm.  First the symmetric positive-definite matrix
$\mathcal N$ is Cholesky-factorized into a symmetric product, $\mathcal
N=LL^\mathrm T$, where $L$ is a lower-triangular matrix.  The generalized
eigenproblem~(\ref{gensym}) is then transformed into an ordinary symmetric
eigenproblem,
       \begin{equation}
\left(L^{-1}\mathcal H L^{-\mathrm T}\right)
\left(L^\mathrm T c\right)
=E\left(L^\mathrm T c\right)
       \end{equation}
The matrix $\left(L^{-1}\mathcal H L^{-\mathrm T}\right)$ is then
diagonalized by any of the available diagonalization methods like the
Jacobi eigenvalue algorithm.

\section{Results}

Given the meson mass, $m_\sigma$, the model has two free parameters:
the strength, $S_\sigma$,
and
the range, $b_\sigma$,
of the coupling operator~(\ref{eq-w-gau}).
It turns out that for any $m_\sigma\in[100,800]$~MeV it is possible to
tune the parameters $S_\sigma$ and $b_\sigma$ such that the binding energy
and the charge radius of the resulting deuteron are reasonable.

For illustration we assume $m_\sigma$=500~MeV similar to
relativistic mean-field models~\cite{manka}.  With $m_\sigma=500$~MeV
and $m_\mathrm{n}=m_\mathrm{p}=939$~MeV the parameters $b_\sigma=3$~fm
and $S_\sigma=20.35$~MeV give the deuteron's ground-state energy
$E_0=-2.2$~MeV and the charge radius $R_c=2.1$~fm which is close to the
accepted values~\cite{codata}.

\begin{figure}
\centerline{\includegraphics{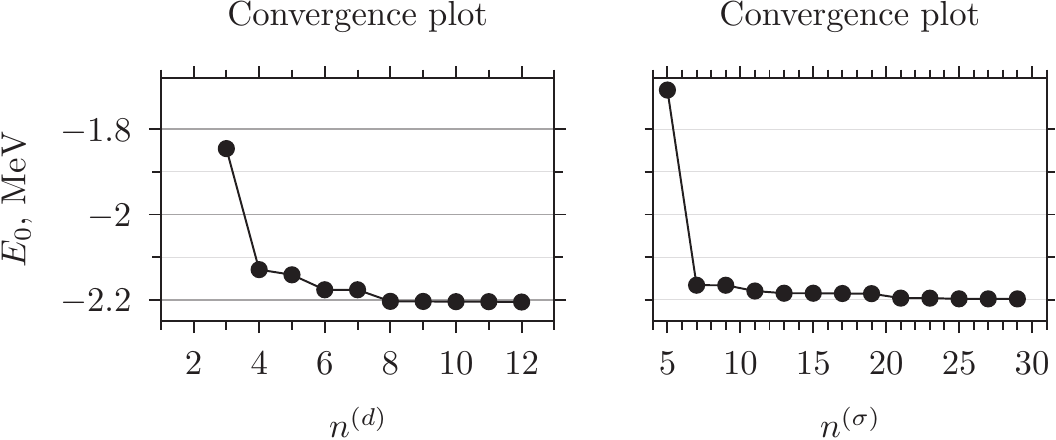}}
\caption{Left:
the deuteron ground-state energy, $E_0$, calculated with
$n^{(d)}$ Gaussians in the two-body subsystem
and
80~Gaussians in the three-body subsystem.
Right:
the deuteron ground-state energy, $E_0$, calculated with
25~Gaussians in the two-body subsystem
and
$n^{(\sigma)}$~Gaussians in the three-body subsystem.
}
	\label{fig-convergence}
	\end{figure}

In this calculation we used low-discrepancy Van der
Corput sampling strategy~\cite{fedorov-quasi}
for the non-linear parameters of the
Gaussians. With this strategy the energy converges
within 1\% with about 10 Gaussians in the two-body subsystem and about
25 Gaussians in the three-body subsystem, as illustrated
in~Fig\ref{fig-convergence}.
The final calculation has been done with
$n^{(d)}=25$
and
$n^{(\sigma)}=80$.

\begin{figure}
\centerline{\includegraphics{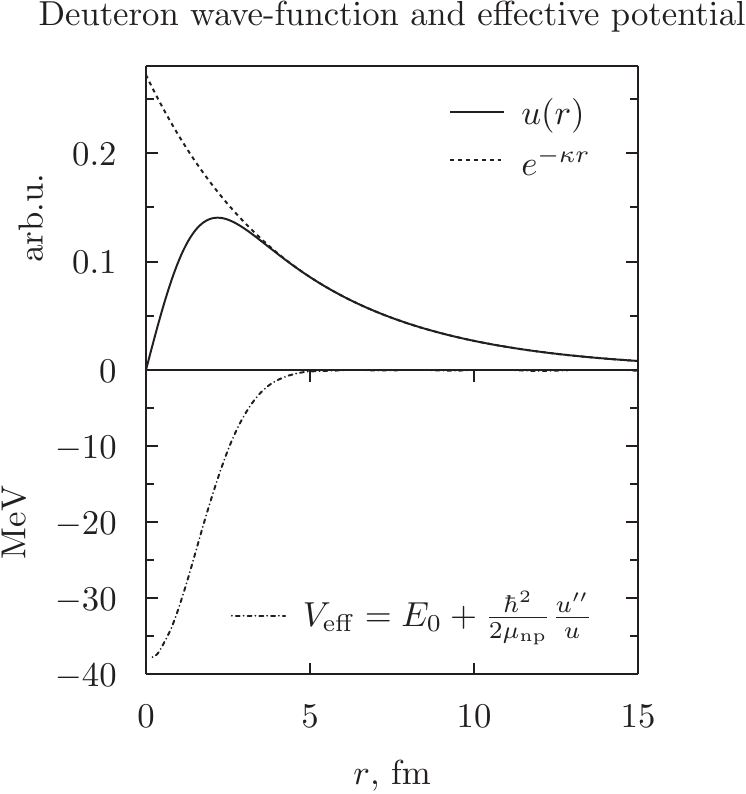}}
\caption{Top: the radial wave-function, $u(r)=r\psi_\mathrm{np}(r)$, of
the neutron-proton subsystem together with its asymptotic
form,~$\exp(-\kappa r)$, where $\kappa=\sqrt{2\mu_\mathrm{np}|E_0|/\hbar^2}\;$;
bottom: the effective potential that produces the same radial
wave-function as in the top figure.}
	\label{fig-psi} \end{figure}

The radial wave-function of the deuteron in the two-body subsystem
is shown in Fig~\ref{fig-psi}.  For
illustration the potential that produces the same wave-function via a
radial Schrodinger equation is also shown.  The potential is short-range
and finite at the origin.

The contribution of the three-body $\sigma$np-subsystem to the norm of
the total wave-function is only about 2\%.  This justifies the assumption
that the two-meson contribution might be a small correction.

\section{Conclusion}
A nuclear model has been introduced where the nucleons interact by
emitting and absorbing mesons which are treated explicitly.  A nucleus
in this model exists in a quantum superposition of states with increasing
number of generated mesons.

The model has been applied to the deuteron---in the model's simplest,
one sigma-meson, incarnation---where the deuteron is a superposition of a
two-body neutron-proton state and a three-body neutron-proton-sigma state.
The model is able to produce a bound deuteron with very reasonable values
of the binding energy and the charge radius.

The anticipated advantages of the model, compared to the phenomenological
potential models, are the reduced number of parameters, natural inclusion
of few-body forces, and natural inclusion of mesonic physics.

The next step in the development of the model could be inclusion of pions.

\section{Appendix}
\subsection{Stochastic sampling of Gaussian parameters}
The Gaussians can be parameterized in the form
	\begin{equation}
\langle \vec r\bigm| A\rangle
=
\exp\left(
-\sum_{i<j=1}^N\left(\frac{\vec r_i-\vec r_j}{b_{ij}}\right)^2
\right)
\equiv
\exp\left(
-\vec r^\mathrm{T} A \vec r
\right)
\;,
	\end{equation}
where $\vec r_i$ is the coordinate of the $i$-th particle and the
matrix $A$ is given as
	\begin{equation}
A=
\sum_{i<j=1}^n
\frac{w_{ij}w_{ij}^\mathrm{T}}{b_{ij}^2}
\;,
	\end{equation}
where the column-vectors $w_{ij}$ are defined through the equation
	\begin{equation}
\vec r_i-\vec r_j=w_{ij}^\mathrm{T}\vec r
\;.
	\end{equation}
In the laboratory frame
$\vec r=\begin{pmatrix}\vec r_1 & \vec r_2 & \dots & \vec r_N\end{pmatrix}^\mathrm{T}$
and the $w_{ij}$ are given for a two-body system as
	\begin{equation}
w_{12}=
\begin{pmatrix}
1 \\ -1
\end{pmatrix}
\;,
	\end{equation}
and for a three-body system as
	\begin{equation}
w_{12}= \begin{pmatrix} 1 \\ -1 \\0 \end{pmatrix}
\;,\;\;
w_{13}= \begin{pmatrix} 1 \\ 0 \\ -1 \end{pmatrix}
\;,\;\;
w_{23}= \begin{pmatrix} 0 \\ 1 \\ -1 \end{pmatrix}
\;.
	\end{equation}
Under a coordinate transformation $\vec r\rightarrow J\vec r$
the column-vectors $w_{ij}$ transform as $w_{ij}\rightarrow
U^\mathrm{T}w_{ij}$ where $U=J^{-1}$.

The range parameters $b_{ij}$ of the Gaussians are chosen stochastically
from the exponential distribution,
	\begin{equation}
b_{ij}=-\ln(u)b
\;,
	\end{equation}
where the quasi-random number $u\in]0,1[$ is taken from a Van der Corput
sequence~\cite{fedorov-quasi} with the scale $b=3$~fm.  Separate sequences
with different prime bases are used for each $ij$-combination.

\subsection{Charge radius}
We define the charge radius, $R_c$, of an $N$-body system as
       \begin{equation}\label{eq-rc}
R_c^2
= \sum_{i=1}^{N} Z_i\langle\vec r_i^2\rangle
= \sum_{i=1}^{N} Z_i\langle\vec r^\mathrm{T}w_iw_i^\mathrm{T}\vec r\rangle
\;,
       \end{equation}
where the summation goes over the bodies in the system; $Z_i$ is the
charge of the body in unit charges; $\vec r_i$ is the coordinate of the
body (in the center-of-mass frame); the brackets $\langle \rangle$~signify
the expectation value in the given state of the system; and
the column-vector $w_i$ is defined via the formula
	\begin{equation}
\vec r_i=w_i^{\mathrm T}\vec r \;.
	\end{equation}
In the laboratory frame, for a two-body system
	\begin{equation}
w_1=\begin{pmatrix} 1 \\ 0 \end{pmatrix}\;,\;\; w_2=\begin{pmatrix} 0 \\
1 \end{pmatrix}\;,\;\;
	\end{equation}
and for a three-body system
	\begin{equation}
w_1= \left( \begin{array}{c} 1 \\ 0 \\ 0 \end{array} \right) \;, w_2=
\left( \begin{array}{c} 0 \\ 1 \\ 0 \end{array} \right) \;, w_3= \left(
\begin{array}{c} 0 \\ 0 \\ 1 \end{array} \right) \;.
	\end{equation}
Under a coordinate transformation $\vec r\rightarrow J\vec r$ the
column-vector $w_i$ transforms with the inverse matrix $U=J^{-1}$ as
$w_i\rightarrow U^{\mathrm T}w_i$  (see subsection~\ref{coo} for a
transformation to the center-of-mass frame).

Now
the matrix element in~(\ref{eq-rc}) between two Gaussians is given
as~\cite{fedorov2017analytic},
       \begin{equation}
\left\langle A\bigm| \vec r^\mathrm T w_iw_i^\mathrm T \vec r \bigm|
A'\right\rangle = \frac32 w_i^\mathrm T(A+A')^{-1}w_i \left\langle
A\bigm|A'\right\rangle \;.
       \end{equation}

\subsection{Coordinate transformations}
\label{coo}
Under a linear coordinate transformation to a new set of coordinates,
	\begin{equation}
\vec r\rightarrow \vec x=J\vec r \;,
	\end{equation}
the matrix elements with correlated
Gaussians preserve their mathematical form as long as the determinant of the
transformation matrix $J$ equals one (otherwise they have to
be divided by the determinant), and as long as one makes the corresponding
transformations of the related matrices and column-vectors:
the kinetic energy matrix transforms as\footnote{which follows
from the identity (with implicit summation notation),
	\begin{equation}
\frac{\partial}{\partial\vec r_i}
K_{ij}
\frac{\partial}{\partial\vec r_j}
=
\frac{\partial}{\partial\vec x_k}
\frac{\partial\vec x_k}{\partial\vec r_i}
K_{ij}
\frac{\partial}{\partial\vec x_l}
\frac{\partial\vec x_l}{\partial\vec r_j}
=
\frac{\partial}{\partial\vec x_k}
J_{ki}K_{ij}J_{lj}
\frac{\partial}{\partial\vec x_l}
\;.
	\end{equation}
}
	\begin{equation}
K\rightarrow JKJ^\mathrm{T} \;,
	\end{equation}
and the $w_i$ (and $w_{ij}$) column-vectors transform as\footnote{which
follows from the identity
	\begin{equation}
w_i^\mathrm{T}\vec r
=
w_i^\mathrm{T}U\vec x
=
(U^\mathrm{T}w_i)^\mathrm{T}\vec x
\;.
	\end{equation}
}
	\begin{equation}
w_i\rightarrow U^\mathrm{T}w_i \;,
	\end{equation}
where $U=J^{-1}$.

One practical set of coordinates are the Jacobi coordinates defined as
	\begin{equation}
\vec x_{i<N} = \frac{\sum_{k=1}^i m_k\vec r_k}{\sum_{k=1}^i m_k}-\vec r_{i+1}\;,\;
\vec x_N = \frac{\sum_{k=1}^N m_k\vec r_k}{\sum_{k=1}^N m_k}\;,
	\end{equation}
where the last coordinate, $\vec x_N$, is the center-of-mass coordinate
that can be omitted if no external forces are acting on the system. This
is equivalent to simply discarding the last row of the $J$ matrix and
the last column of the $U$ matrix.

\bibliographystyle{hunsrt}
\bibliography{ms}

\begin{thebibliography}{10}

\bibitem{yukawa}
H.~Yukawa.
\newblock On the interaction of elementary particles.
\newblock {\em Proc. Phys. Math. Soc. Jap.}, 17:48, 1935.

\bibitem{jensen}
Phillip~J. Siemens and Aksel~S. Jensen.
\newblock {\em Elements Of Nuclei: Many-body Physics With The Strong
  Interaction}.
\newblock Addison-Wesley, 1987.

\bibitem{langanke}
R.~Machleidt.
\newblock One-boson-exchange potentials and nucleon-nucleon scattering.
\newblock In K.~Langanke, J.A. Maruhn, and S.E. Koonin, editors, {\em
  Computational Nuclear Physics 2 Nuclear Reactions}, pages 1--29.
  Springer-Verlag, 1993.

\bibitem{bonn}
R.~{Machleid}, K.~Holinde, and Ch. Elster.
\newblock The {B}onn meson-exchange model for the nucleon—nucleon
  interaction.
\newblock {\em Physics Reports}, 149(1):1--89, 1987.

\bibitem{av18}
R.B. Wiringa, V.G.J. Stoks, and R.~Schiavilla.
\newblock Accurate nucleon-nucleon potential with charge-independence breaking.
\newblock {\em Phys. Rev. C}, 51(1):38, 1995, arXiv:nucl-th/9408016.

\bibitem{historical}
R.~Machleidt.
\newblock Historical perspective and future prospects for nuclear interactions.
\newblock {\em Int. J. Mod. Phys. E}, 26:1730005, 2017, arXiv:1710.07215
  [nucl-th].

\bibitem{sigma-meson}
M.~Albaladejo and J.~A. Oller.
\newblock Size of the $\sigma$ meson and its nature.
\newblock {\em Phys. Rev. D}, 86:034003, 2012, arXiv:1205.6606 [hep-ph].

\bibitem{manka}
R.~{Manka} and I.~{Bednarek}.
\newblock Nucleon and meson effective masses in the relativistic mean-field
  theory.
\newblock {\em J. Phys. G: Nucl. Part. Phys.}, 27(10):1975, 2001,
  arXiv:nucl-th/0011084v2.

\bibitem{suzuki-varga}
Yasuyuki Suzuki and Kálmán Varga.
\newblock {\em Stochastic Variational Approach to Quantum-Mechanical Few-Body
  Problems}.
\newblock Springer, 1998.

\bibitem{fedorov2017analytic}
D.V. Fedorov.
\newblock Analytic matrix elements and gradients with shifted correlated
  gaussians.
\newblock {\em Few-Body Systems}, 58(1):21, 2017, arXiv:1702.06784 [nucl-th].

\bibitem{codata}
Peter~J. Mohr, David~B. Newell, and Barry~N. Taylor.
\newblock {CODATA} recommended values of the fundamental physical constants:
  2014.
\newblock arXiv:1507.07956 [physics.atom-ph].

\bibitem{fedorov-quasi}
D.V. Fedorov.
\newblock Correlated gaussians and low-discrepancy sequences.
\newblock {\em Few-Body Systems}, 60(3):55, 2019, arXiv:1910.05223
  [physics.comp-ph].

\end{thebibliography}
\end{document}